\pdfoutput=1
\documentclass[12pt,a4paper]{article}

\usepackage{amsfonts,amsthm}
\usepackage{jcappub}

\title{Plateau Inflation from Random Non-Minimal Coupling}

\author[1]{Benedict J. Broy}
\author[2,3]{Dries Coone}
\author[2]{Diederik Roest}

\affiliation[1]{\emph{Deutsches Elektronen-Synchrotron DESY, Theory Group, 22603 Hamburg, Germany}}
\affiliation[2]{\emph{Van Swinderen Institute for Particle Physics and Gravity, University of Groningen, Nijenborgh 4, 9747 AG Groningen, The Netherlands}}
\affiliation[3]{\emph{Theoretische Natuurkunde, Vrije Universiteit Brussel and The International Solvay Institutes, Pleinlaan 2, B-1050 Brussels, Belgium}}

\emailAdd{benedict.broy@desy.de}
\emailAdd{a.a.coone@rug.nl}
\emailAdd{d.roest@rug.nl}

\abstract{A generic non-minimal coupling can push any higher-order terms of the scalar potential sufficiently far out in field space to yield observationally viable plateau inflation. We provide analytic and numerical evidence that this generically happens for a non-minimal coupling strength $\xi$ of the order $N^2_e$. In this regime, the non-minimally coupled field is sub-Planckian during inflation and is thus protected from most higher-order terms. For larger values of $\xi$, the inflationary predictions converge towards the sweet spot of PLANCK. The latter includes $\xi \simeq 10^4$ obtained from CMB normalization arguments, thus providing a natural explanation for the inflationary observables measured.}

\arxivnumber{1604.05326}

\begin{document}

\maketitle
\flushbottom

\section{Introduction}

The theory of cosmic inflation \cite{Linde:1983gd, Starobinsky:1980te} has become the leading paradigm to explain the initial conditions of the early universe. Combined with cosmological perturbation theory \cite{Mukhanov:1990me}, it provides a mechanism for seeding structure formation that is in astonishing agreement with recent observations \cite{Bennett:2012zja, Ade:2015lrj, Ade:2015tva}. 

Inflation is typically taken to arise from the potential energy of a scalar field mimicking the behaviour and equation of state of a cosmological constant $P_\Lambda=-\rho_\Lambda$. However, radiative corrections to the inflaton mass or generically higher dimensional operators may spoil the required flatness of the inflaton potential. Specifically, the slow-roll parameter $\eta$ may receive corrections of order one and subsequently observationally viable slow-roll inflation is no longer possible; this is referred to as the $\eta$-problem. In order to circumvent this problem, one can invoke an approximate continuous shift symmetry $\chi\rightarrow\chi+ const.$ of the inflaton $\chi$. 

An alternative method to ensure the flattening of the scalar potential is by introducing a non-minimal coupling of the inflaton field to gravity. Following previous works \cite{Salopek:1988qh, Bezrukov:2007ep,Kallosh:2013tua}, we consider a Jordan frame in which the non-minimal coupling has a polynomial expansion around the minimum of the potential energy. In this case, a generic polynomial expansion of the non-minimal coupling and the potential energy results in a shift-symmetric Einstein frame that is protected from corrections by the non-minimal coupling strength $\xi$. This ansatz ensures, at least for intermediate fields, the existence of an approximate shift symmetry which then may serve to drive an inflationary phase.

By including the non-minimal coupling, we extend the previous approach to define arbitrary polynomials for the potential (or, equivalently in slow roll inflation, the Hubble function) originating from the Hubble flow code by Kinney \cite{Kinney:2002qn}. In this minimally coupled case, this approach confirmed the prediction from Hoffman and Turner in \cite{Hoffman:2000ue} that a polynomial Hubble function results in the generic shape 
 \begin{align}
  3r=16(1-n_s) \,, \quad \text{or} \quad r=0 \,. \label{min-predictions}
 \end{align} 
However, this generic result is defined by the polynomial hypothesis \cite{Ramirez:2005cy,Chongchitnan:2005pf,Vennin:2014xta} and the assumption of minimal coupling to gravity, as we will show below. Moreover, these generic predictions are not consistent with the observations from PLANCK \cite{Ade:2015lrj}. 

In the current paper, we extend these investigations by including a non-minimal coupling 
 \begin{align}
  \Omega(\phi) = 1 + \xi f(\phi) \,, \label{nonminimal}
 \end{align} 
that contains an arbitrary polynomial $f$. The main physical parameter in our theory is the strength of the non-minimal coupling $\xi$. It turns out that one can identify a number of distinct regimes for this parameter. A number of these were outlined in  \cite{Kallosh:2013tua} for the simple case that the scalar potential and the non-minimal coupling are related by a square relation as 
 \begin{align}
  V_J = \lambda f(\phi)^2 \,. \label{square}
 \end{align}
These will be recapped in section 2. 

In this paper we focus on the more general case where the scalar potential and the non-minimal coupling are given by different and arbitrary polynomials $f$ and $g$:
 \begin{align}
  \Omega=1+\xi f(\phi)\thinspace,\quad V_J = \lambda\thinspace g(\phi) \thinspace . \label{general1}
 \end{align}
In this case the large-field plateau can be destroyed by the different field dependence of both functions. The point where this happens depends on the non-minimal coupling strength; we identify the following two regimes, where $N_e$ denotes the number of e-folds at horizon exit of the scales now observable through the CMB: at $\xi \sim N_e^2$ there is a universal form for the inflationary predictions which converge for $\xi > N_e^2$ to those identical to the Starobinsky model. We will provide evidence for these two stages both from analytical expressions as well as from numerical investigations; a first illustration can be seen in figure 1. Our study thus lends further support to pinpoint the non-minimal coupling strength to $\xi\simeq 10^4$, following the argument from the scalar amplitude normalization and the toy model discussion of \cite{Broy:2014sia}.

\begin{figure}[t!]
\begin{center}
\includegraphics[scale=0.6]{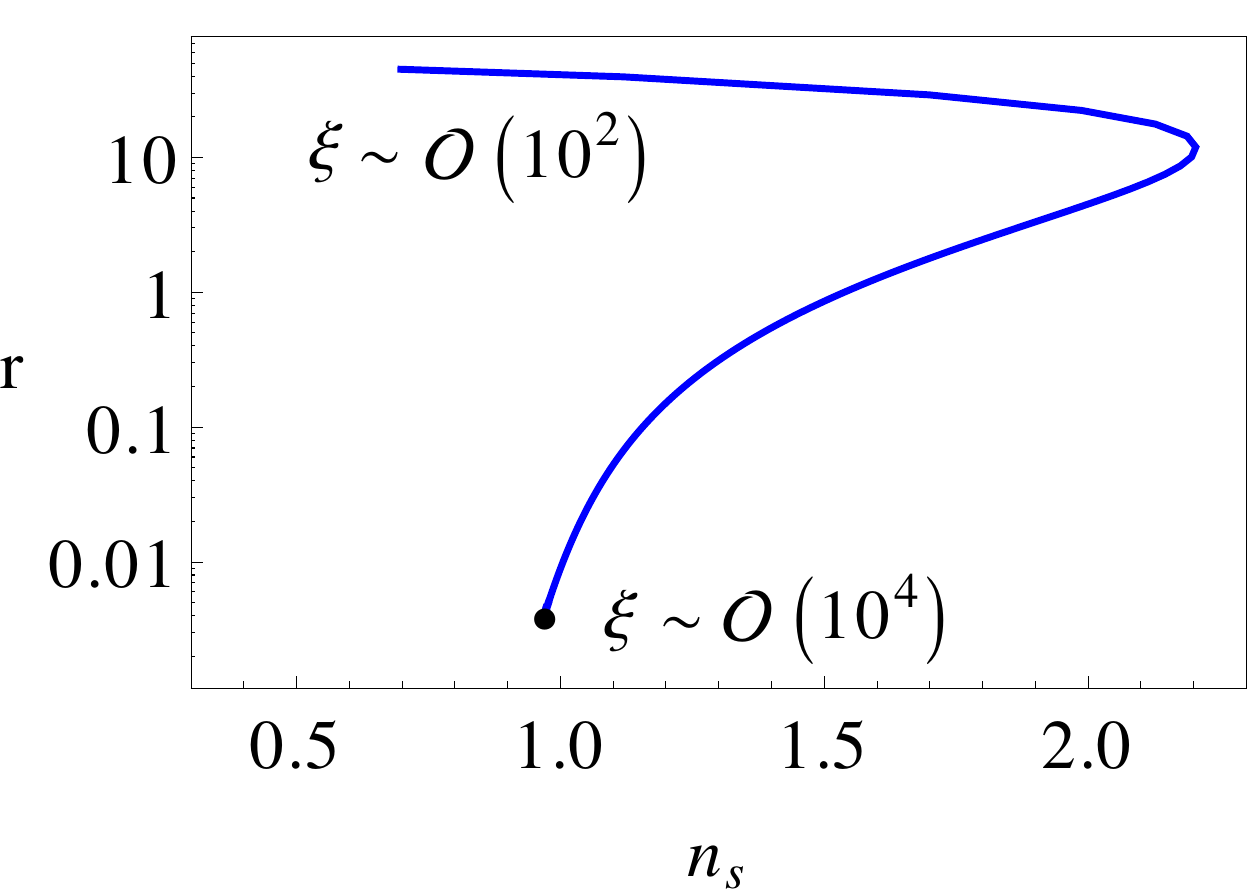} \includegraphics[scale=0.6]{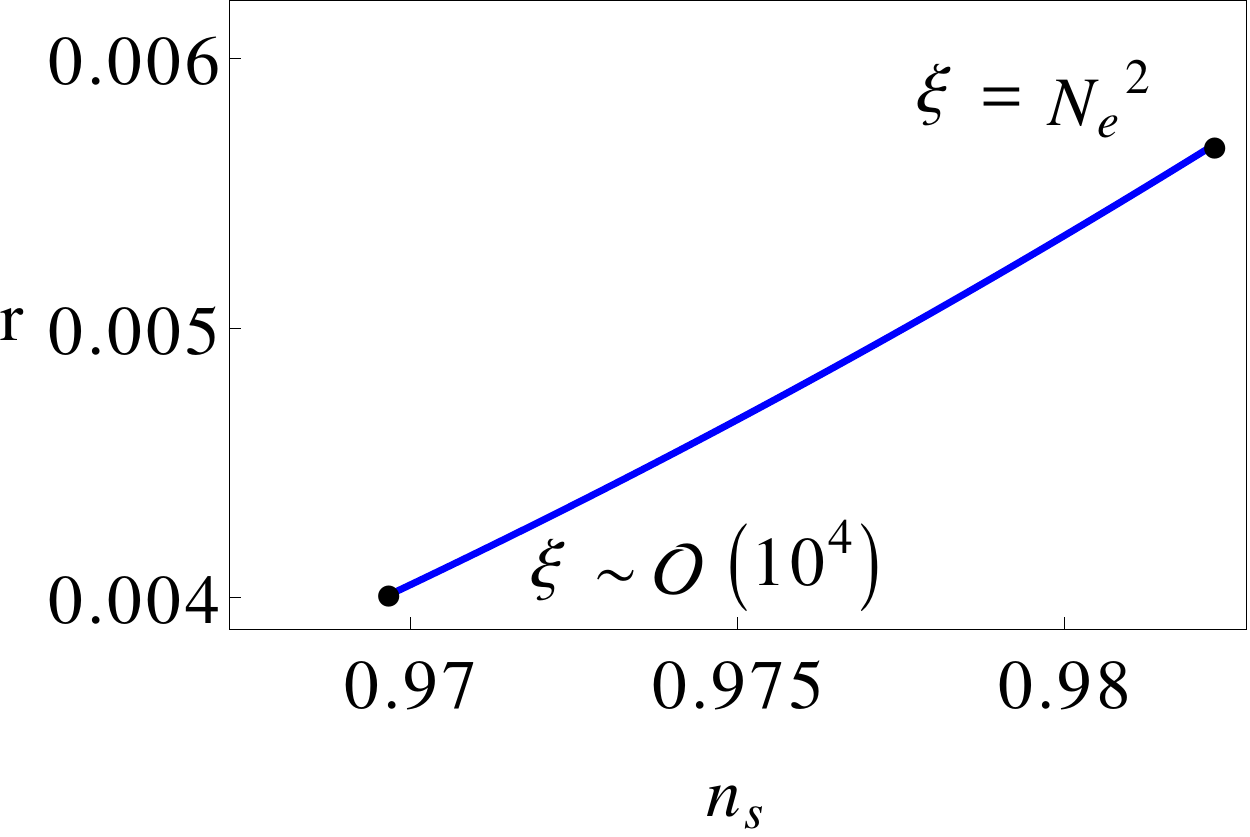}
\caption{\textbf{Left:} \emph{The $n_s, r$ predictions for gradually increased non-minimal coupling $\xi$ at $N_e=55$ of an example with $\Omega = 1 + \xi \phi$ and $V_J=e^\phi-1-\phi$. Note that for lower $\xi$, the predictions are clearly incompatible with observational bounds. }\textbf{Right:} \emph{Predictions of the same model for $N_e^2\leq\xi\lesssim \mathcal O(10^4)$ with $N_e=55$. Increasing $\xi$ to values $\xi\gtrsim\mathcal O(10^4)$ has all further data points precisely cluster at the sweet spot of PLANCK. }}
\label{trajectory}
\end{center}
\end{figure}


The rest of the paper is structured as follows. We start with a short review of the universal attractor. We continue to generalize this set-up to arbitrary non-minimal coupling functions and potentials and demonstrate how the coupling strength $\xi$ may ensure a sufficient amount of observationally viable inflation. After outlining the analytic approximate expressions for the inflationary observables, we employ numerical methods to scan the landscape of possible inflationary scenarios with arbitrary coefficients. We conclude in the discussion and outline further analytical and numerical evidence in the appendix.

\section{Non-minimally coupled inflation} 

We start with a brief recollection of the universal attractor \cite{Kallosh:2013tua}, which  may be seen as a generalization of Higgs inflation \cite{Salopek:1988qh, Bezrukov:2007ep}. Consider the Jordan frame Lagrangian
\begin{equation}\label{JordanL}
\frac{\mathcal L_J}{\sqrt{-g_J}}=\frac{1}{2}\Omega(\phi)R_J-\frac{1}{2}\left(\partial\phi\right)^2-V_J(\phi),
\end{equation}
with non-minimal coupling \eqref{nonminimal} and scalar potential \eqref{square}. Going to the Einstein frame via
\begin{equation}
g^E_{\mu\nu}=\Omega(\phi)g^J_{\mu\nu},
\end{equation}
where the superscripts denote Einstein and Jordan frame respectively, the Lagrangian becomes
\begin{equation}\label{EinsteinL}
\frac{\mathcal L_E}{\sqrt{-g_E}}=\frac{1}{2}R_E-\frac{1}{2}\left[\frac{1}{\Omega}+\frac{3}{2}\left(\frac{\partial \ln \Omega}{\partial\phi} \right)^2 \right]\left(\partial\phi \right)^2 -\frac{V_J}{\Omega^2}. 
\end{equation}
As a function of the coupling strength $\xi$, the main features of this inflationary model are \cite{Kallosh:2013tua}:
\begin{itemize}
\item
 $\xi = 0$: The minimally coupled case with a random scalar potential yields inflationary predictions $n_s^{(0)}$ and $r^{(0)}$ that interpolate between small-field plateau  and large-field chaotic inflation  \eqref{min-predictions} \cite{Ramirez:2005cy,Chongchitnan:2005pf,Vennin:2014xta}. Almost all of these are ruled out by the PLANCK results.
\item
Very small $\xi$: At weak coupling, there is a universal behavior for the inflationary predictions. Retaining only linear terms in the coupling strength $\xi$ one finds \cite{Kallosh:2013tua}
 \begin{align}
  n_s = n_s^{(0)} + \tfrac{1}{16} \xi f r^{(0)} , \quad r = r^{(0)} - \xi f r^{(0)} \,.
 \end{align}
Note that the inflationary predictions therefore have the same behaviour in the $(n_s,r)$ plane, corresponding to a downward line with a slope of $-16$.
 \item
  Finite $\xi< \mathcal{O}(1)$: The original behaviour will be flattened at large field values that are beyond the region probed by the cosmic microwave background (CMB); horizon exit of CMB scales takes place closer to the minimum and hence allows for a wide range of inflationary predictions depending on the specifics of the polynomial potential. In particular, in this regime one looses the simplicity of the linear approximation, resulting in a wide range of different behaviours. 

For Higgs inflation, this regime is a particularly simple straight line, again with a slope of $-16$, that interpolates between quartic and Starobinsky inflation; for other starting points, the results of this regime are very different and generically complicated.  
 \item 
 Finite $\xi \gtrsim \mathcal{O}(1)$: Increasing the non-minimal coupling to and beyond order-one values pushes the plateau sufficiently close to the minimum of the scalar potential, yielding predictions that are indistinguishable from Starobinsky inflation:
\begin{align} \label{universal}
  n_s = 1 - \frac{2}{N_e}+\frac{3}{2}\frac{\log(N_e)}{N_e^2} + \ldots \thinspace ,\quad r= \frac{12}{N_e^2} -18\frac{\log(N_e)}{N_e^3} +\ldots
\end{align}
where $N_e$ denotes the number of e-folds before the end of inflation and we have included subleading corrections from \cite{Roest:2013fha} to the well known leading order result. The exact value of $\xi$ where this happens depends on the specific choice of scalar potential. A derivation of expressions \eqref{universal} (and the later given \eqref{spectralindex}) will be provided in appendix \ref{derivation}.
\end{itemize}
The simplification of the latter limit arise as the first term in the kinetic function is sufficiently suppressed:
\begin{equation}\label{UArelations}
\Omega\ll \frac{3}{2}\Omega\thinspace'^{\thinspace 2} \,.
\end{equation}
In terms of a canonically normalized scalar field $\chi$,
\begin{equation}\label{omega}
\Omega(\chi)=e^{\sqrt{2/3}\chi},
\end{equation}
the scalar potential becomes
\begin{equation}\label{staro}
V_E=\frac{\lambda}{\xi^2}\left(1-e^{-\sqrt{2/3}\thinspace \chi} \right)^2.
\end{equation}
This is conformally dual to $R^2$-inflation \cite{Starobinsky:1980te, Whitt:1984pd}, and results in the relation
 \begin{align}
  N_e\sim\frac{3}{4}(\Omega-1) \,, \label{N}
 \end{align}
for the number of e-folds.

Already in the original paper \cite{Kallosh:2013tua} it was argued that taking an independent scalar potential \eqref{general1} does not change the leading inflationary predictions as long as the function $g(\phi)$ and the square of $f(\phi)$ share the order of their first zero while the non-minimal coupling is taken sufficiently strong. A first quantitative investigation for a toy model of higher order corrections demonstrated that the leading order behaviour of the universal attractor can indeed be made robust once a certain value of the non-minimal coupling $\xi$ is chosen \cite{Broy:2014sia}. Specifically, the Jordan frame potential was taken to be a function of the non-minimal coupling $f(\phi)$, i.e.\
\begin{equation}
V_J(\phi)\rightarrow V_J(f(\phi)).
\end{equation}
This allowed the function $f(\phi)$  to be left completely unspecified. The deviation of $V_J(f)$ from a quadratic function was then used to model corrections to the universal attractor behaviour. Different types of expansions with $\mathcal O(1)$ coefficients were employed, from simple monomials to different series. Remarkably, it was found that a coupling strength of $\xi\sim\mathcal O(10^4)$ was sufficient to maintain the leading order inflationary predictions.  

The observation that a  sufficiently large $\xi$ can, regardless of an \emph{infinite} tower of higher order corrections with order one coefficients, induce a Starobinsky-like inflationary plateau over a finite field range derives from $\xi$ being able to drive $\Delta\phi<1$ when increased. Hence all higher order terms in the Jordan frame potential are sub-leading. In other words, the effect of higher order terms can simply be pushed far away in canonical field space by sufficiently enlarging the non-minimal coupling strength $\xi$. 

The above study was however not conducted with arbitrary coefficients and an expansion of the scalar potential in terms of the non-minimal coupling might not be the most generic. In this work, we aim to study arbitrary corrections with a more generic ansatz and hence to find how to generically alleviate the $\eta$-problem of arbitrary potentials.

\section{Analytic predictions}\label{retain}

The aim of this section is to explicitly show the robustness of the inflationary potential from an arbitrary number of higher order terms. Consider the non-minimal coupling or frame function as well as the potential to be arbitrary polynomials with the only requirement that the Jordan frame potential and the square of the frame function share the order of their first zero for $\phi$; in particular, we require the Jordan frame potential to have a minimum and the frame function to contain a term linear in the Jordan frame field $\phi$. We thus make the following ansatz
\begin{equation}
\label{general}
  \Omega(\phi)=1+\xi \sum_{n=1}^{M_\Omega} a_n\phi^n,\quad V_J(\phi)=\lambda \sum_{m=2}^{M_V}b_m\phi^m \,,
\end{equation}
where
 \begin{itemize}
 \item 
 We have kept the factor $\lambda$ to be consistent with the original work and will assume it to take a natural value of $\lesssim\mathcal O(1)$. 
 \item 
We assume $b_2$ and $a_1$ to be positive in order to ensure a Minkowski minimum at $\phi =0$ and that $\chi$ and $\phi$ both decrease at the same time ($d\phi/d\chi>0$) close to the minimum. 
 \item
 We have introduced $M_{\Omega, V}$ to denote the respective cut-off of both series. These will not play a role in the analytic part; in principle, both polynomials may contain an infinite number of terms.
\end{itemize}
For the general set-up \eqref{general}, and for now assuming to be in the regime $\phi<1$, the expression for the number of e-folds of \eqref{N} obtains corrections as
\begin{equation}\label{importantN}
N_e\sim\frac{3}{4}\thinspace\Omega-\frac{b_3 \Omega^3}{8\thinspace b_2a_1 \xi} + \mathcal O^{(2)}\left(\frac{\Omega^2}{\xi}\right),
\end{equation}
which may be understood as an expansion in $\Omega^2/\xi$. From the zeroth-order relation \eqref{N} for the number of e-folds, we find that the lower bound on the non-minimal coupling strength for generating a sufficient amount of inflation within $\Delta\phi<1$ is 
\begin{equation}\label{start}
\xi  \gtrsim \mathcal O(N^2_e).
\end{equation}
We will assume this in what follows.

To obtain a value for $\xi$ that ensures the corrections to be sufficiently far away from the inflaton's minimum and to have inflation matching observations by PLANCK, it is most useful to study the inflationary observables and their dependence on the infinite tower of higher order terms.
To leading order, the expressions for the inflationary observables $n_s$ and $r$ of \eqref{general} are given by 
\begin{align}\label{spectralindex}\notag
n_s=&1-\frac{2}{N_e}+\frac{64}{27}\frac{b_3}{b_2} \left(\frac{N_e}{a_1\xi} \right)+{\mathcal O}^{(2)}\left( \frac{1}{N_e}, \frac{N_e}{a_1\xi} \right),\\ 
r=&\frac{12}{N_e^2}+\frac{128}{9}\frac{b_3}{b_2}\left(\frac{1}{a_1\xi}\right)+ {\mathcal O}^{(2)}  \left(\frac{N_e}{a_1\xi} \right) \notag \\
 &+ {\mathcal O}^{(3)}\left( \frac{1}{N_e}\right)\thinspace ,
\end{align}
which is in line with \cite{Broy:2015qna}.\footnote{For a more detailed derivation, please see appendix \ref{derivation}.} Expressions \eqref{spectralindex} are expansions in $1/N_e$ and $N_e/(a_1\xi)$. For the spectral index $n_s$, the leading order terms are the linear contributions of the $1/N_e$ and the $N_e/(a_1\xi)$ expansions. For the tensor to scalar ratio $r$, the leading order terms are the quadratic and bilinear expressions of both expansions (note that we only give two of these three terms). Further subleading terms stem from higher order and cross terms in $1/(a_1\xi)$ and $N_e/(a_1\xi)$ and are denoted by ${\mathcal O}^{(n)}$. Note that we have omitted the subleading corrections of \cite{Roest:2013fha}, i.e.\ higher order terms in $\log(N_e)/N_e$, for clarity. 

For $n_s$ and $r$ to be dominated respectively by the linear and quadratic term in $1/N_e$, i.e.\ for prolonging the intermediate plateau of the Einstein frame potential, we quickly identify the requirement \eqref{start}, self-consistent with the derivation's starting point. This hence marks the onset of a convergence of the inflationary predictions towards the values measured. Moreover, the next to leading order terms come with the same $a_1, b_2, b_3$ dependence. This implies that the ratio of the next to leading order terms has a universal form
\begin{equation}
\frac{\delta r}{\delta n_s}=\frac{6}{N_e} \,. \label{slope}
\end{equation}
This predicts that in the vicinity of the Starobinsky point in an $n_s/r$ scatter plot, there will be deviations to both the bottom left and the top right with a fixed slope that is independent of the specific coefficients. The former of these have $b_3$ negative (note that $a_1$ and $b_2$ have to be positive to guaranty the positivity of the frame function and the potential around the minimum); these corrections induce a hilltop-like deformation to the plateau. Similarly, the predictions to the top right of Starobinsky arise from positive $b_3$ corrections, corresponding to an upward curve in the plateau.

Thus we conclude that in the presence of a {\it generic} non-minimal coupling, to be contrasted to the simpler case satisfying square relation \eqref{square}, we expect the approach to the universal attractor to take place at a later stage (i.e.~larger value of $\xi$) but also in a cleaner manner (i.e.~in a straight line). This is nicely confirmed by figure \ref{fig:nsr_O1}.

Turning to the comparison with observations, for higher order terms not to spoil the value of $n_s$ observed by PLANCK, we consider the 2-$\sigma$ bound by PLANCK of $\delta n_s<0.008$ at $N_e=55$ and find, given $a_1, b_2, b_3 \sim\mathcal O(1)$,
\begin{equation}
\xi\gtrsim  10^4 \,.
\end{equation}
This hence sets, given order one coefficients, a lower bound on the non-minimal coupling strength $\xi$ to realize observationally viable slow-roll inflation. Remarkably, the value of $\xi$ obtained from the requirement of matching the observed spectral index $n_s$ is also similar to the value needed to match COBE normalization\footnote{Recalling $A_s=(24\pi^2)^{-1} V/\epsilon\sim 10^{-9}$ stemming from the CMB temperature data, it readily follows that $\xi\sim 10^5\sqrt{\lambda}$.} (provided the self-coupling $\lambda$ is sub-Planckian).  Thus two independent observational indications -- in technical terms the spectral index $n_s$ and the amplitude $A_s$ -- hint towards an otherwise ad hoc value of the theory's parameter. The length and the height of the inflationary plateau are correctly set by the single parameter $\xi$.

The results of \cite{Broy:2014sia} hence nicely carry over to our more general ansatz \eqref{general}: given a scalar field with a minimum and polynomial non-minimal coupling with strength $\xi\gtrsim 10^4$ as required by the COBE normalization and expressions \eqref{spectralindex}, plateau inflation with PLANCK-like observables will be realised.

\section{Numerical results}

We now turn to the numerical body of this work and study the behaviour of ansatz \eqref{general} given arbitrary coefficients. By choosing random values for $a_n, b_m$, a Monte Carlo analysis can be performed using a procedure based on \cite{Hoffman:2000ue, Kinney:2002qn, Ramirez:2005cy, Coone:2015fha}. The prior distribution for $a_n$ and $b_m$ is chosen to be between $[-1/n!,1/n!]$ in order to represent a Taylor series with an increasing convergence range for large truncation order.\footnote{We will comment on the omission of the factorial suppression in section \ref{discussion}.}

Our numerical model closely follows the approach from \cite{Hoffman:2000ue,Kinney:2002qn, Coone:2015fha}, with some modifications to incorporate the non-canonical kinetic term. Thus sampling the current model in the Einstein frame \eqref{EinsteinL}, but without utilizing the canonical normalization \eqref{omega}. With a non-canonical kinetic term the first two slow-roll parameters become
\begin{align}
\epsilon=\frac{1}{2\thinspace K} \left(\frac{1}{V_J}\frac{\partial V_J}{\partial \phi}-\frac{2}{\Omega}\frac{\partial \Omega}{\partial\phi} \right)^2\thinspace,\quad \eta= \frac{\Omega^2}{K\thinspace V_J}\left[\frac{\partial^2}{\partial\phi^2}\left(\frac{V_J}{\Omega^2}\right)-\frac{1}{2\thinspace K}\frac{\partial K}{\partial\phi}\frac{\partial}{\partial\phi}\left(\frac{V_J}{\Omega^2}\right) \right] \,, \label{def_epsi_g} 
\end{align}
in terms of the non-canonical kinetic function
\begin{equation}
K=\frac{1}{\Omega}+\frac{3}{2}\left(\frac{1}{\Omega}\frac{\partial\Omega}{\partial\phi}\right)^2 \,.
\end{equation}
The number of e-folds then follows as
\begin{align}
N_e=\int\frac{1}{\sqrt{2\epsilon}}d\chi=\int\frac{\sqrt{K}}{\sqrt{2\epsilon}}d\phi \,,
\end{align}
where $\chi$ is the canonical Einstein frame and $\phi$ the non-canonical Jordan frame inflaton. 
Using these expressions for the slow-roll parameters, the rest of the procedure is similar to the ones in \cite{Kinney:2002qn, Coone:2015fha} and is summarized below
\begin{itemize}
	\item Draw parameters $a_n$ and $b_m$ from Eq.~(\ref{general}) according to a uniform distribution.
	\item Calculate $\epsilon, \eta$ from expressions (\ref{def_epsi_g}).
	\item Find the type of the resulting inflationary model (the types will be defined below).
	\item In case inflation ends with $\epsilon=1$ and contains $50$ e-folds, calculate $n_s$ and $r$ using $n_s=1+2\eta-6\epsilon$, $r=16\epsilon$.
\end{itemize}
This procedure is iterated $10^6$ times in all ensembles shown. Note that we are expanding $n_s$ and $r$ only to first order in slow roll, while the accuracy of the figures will imply that we need higher precision. We do not add higher order terms since our goal is to see the approach towards the general attractor, and not to obtain very precise high order predictions for $n_s$ and $r$ in the attractor phase. Moreover, at this moment there is no need to use higher orders of slow roll, since the PLANCK bounds on $n_s$ and $r$ are not precise enough. However, the linear terms in the $1/N_e$ expansion of Eq.~(\ref{universal}) will not be enough in comparison with the numerical data, and in principle higher order terms have to be included to match the accuracy in the figures. Performing this analysis we obtain that the so-called `Starobinsky point' will be at $n_s=0.96157, r=0.004192$ for $N_e=50$ to first order in slow roll.
\begin{figure}[t!]
\centering
\includegraphics[width=0.5\textwidth]{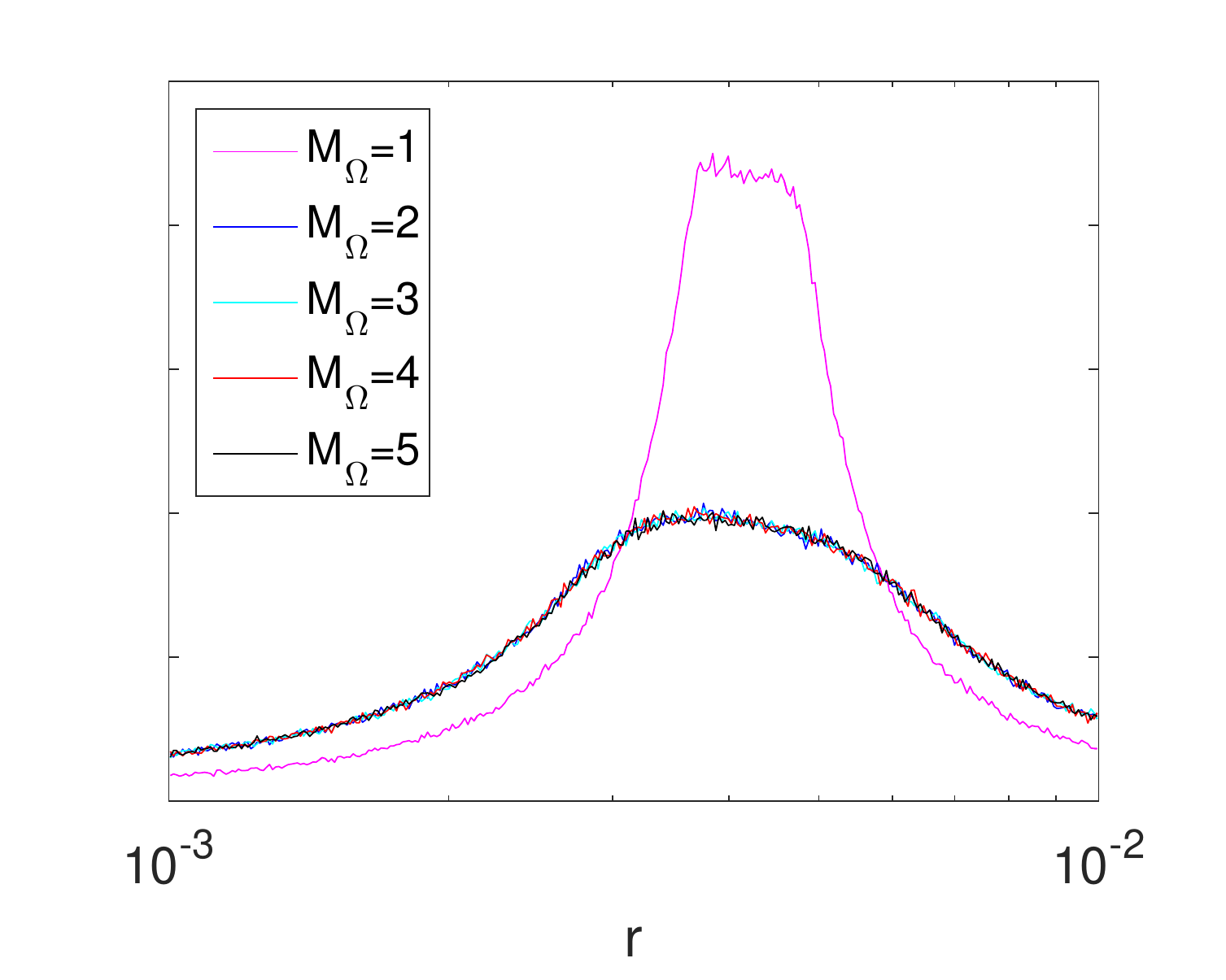}
\caption{\emph{An $r$ density plot, on a linear scale, for different values of $M_\Omega$ with $M_V=10$ and $\xi=10^4$. For $M_\Omega>2$ the system is truncation independent.}}
\label{fig:truncation}
\end{figure}
One should distinguish different late-time behaviours:
\begin{itemize}
\item
The one we are looking for is when $\epsilon$ becomes 1, and then increases to infinity when $\phi\to0$, which we will call a \emph{non-trivial} ending, following the terminology of \cite{Kinney:2002qn}. 
\item
In addition there is the possibility that the model does have an inflation phase with $\epsilon=1$ at the end, but does not include the required $50$ e-folds of inflation. Those models are referred to as \emph{insuf}. 
\item
Besides the non-trivial and insuf endings, there is a fraction of the configurations with a zero in $\Omega$ or $V$ (or both) before inflation starts. Negative potential and frame function are not allowed during inflation, thus we give them the label \emph{$\Omega, V$-negative}.
\item
Finally, a very small fraction of the models does not include an inflation phase at all, but this fraction is negligibly small for the values of $\xi$ discussed in this work.
\end{itemize}
In what follows, we will focus on the non-trivial trajectories.

Secondly, one should worry about the effects of the truncation of the polynomials in \eqref{general}: do the resulting predictions depend on these? Fortunately, at the large $\xi$ values that we are presently interested in, it is computationally possible to include a sufficient number of terms in both the non-minimal coupling and the scalar potential to render our results truncation independent. This is illustrated in Fig.~\ref{fig:truncation}. In what follows, we will consider the specific case of $M_V = 10$ and $M_\Omega = 5$, but none of our results depend on these specific numbers.

\begin{figure}[t!]
\centering
\includegraphics[width=0.5\textwidth]{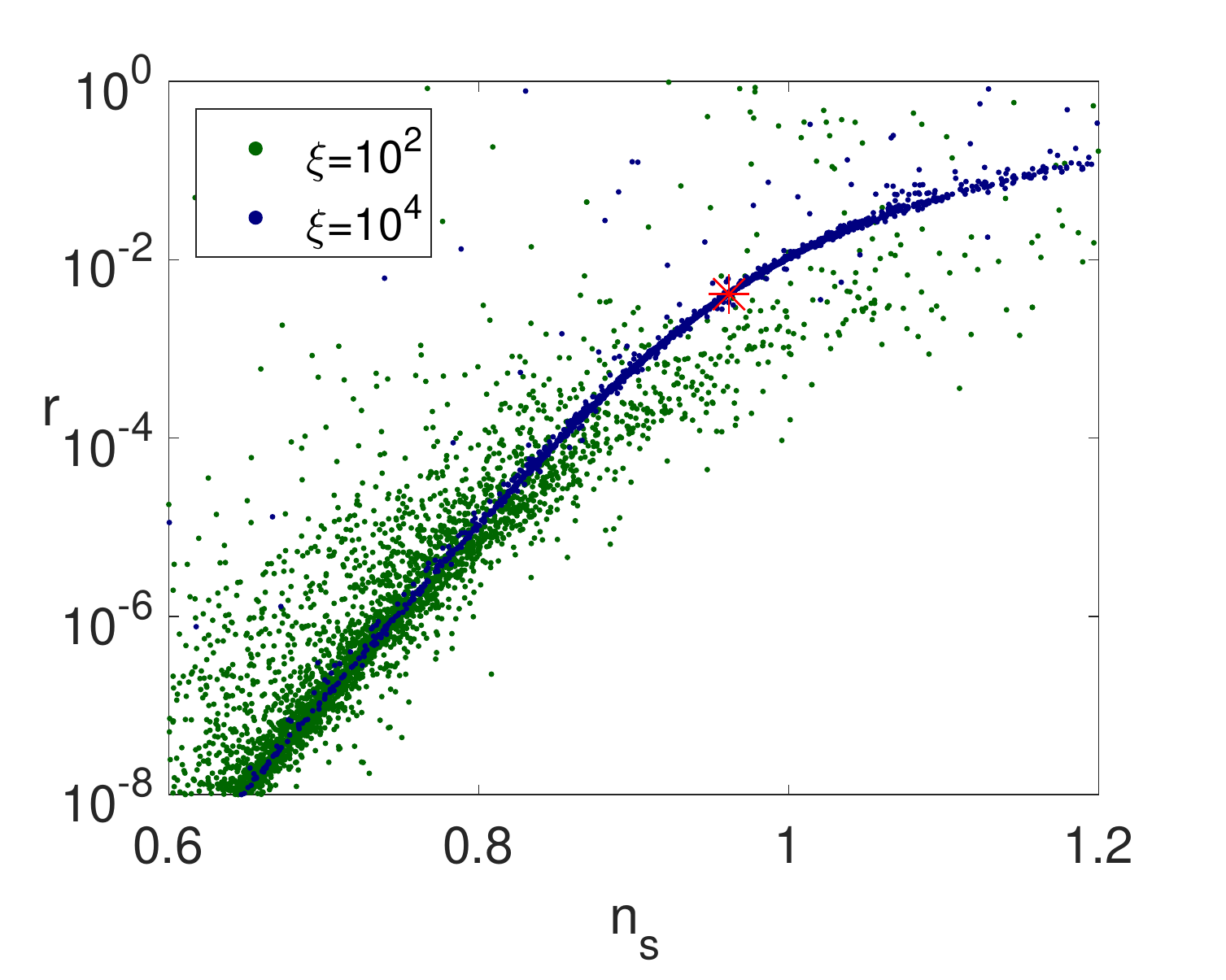}
\caption{\emph{The plot shows a scatter plot of $5000$ trajectories from the ensembles with $M_\Omega=5$ and $M_V=10$ for $\xi=10^2$ in green and $\xi=10^4$ in blue. The $\xi=10^2$ points overlap the $\xi=10^4$ points. The red star represents the Starobinsky point $n_s\approx 0.962, r\approx 0.004$.}}
\label{fig:nsr_O1}
\end{figure}

\begin{figure*}[t!]
\begin{center}
\includegraphics[width=0.48\textwidth]{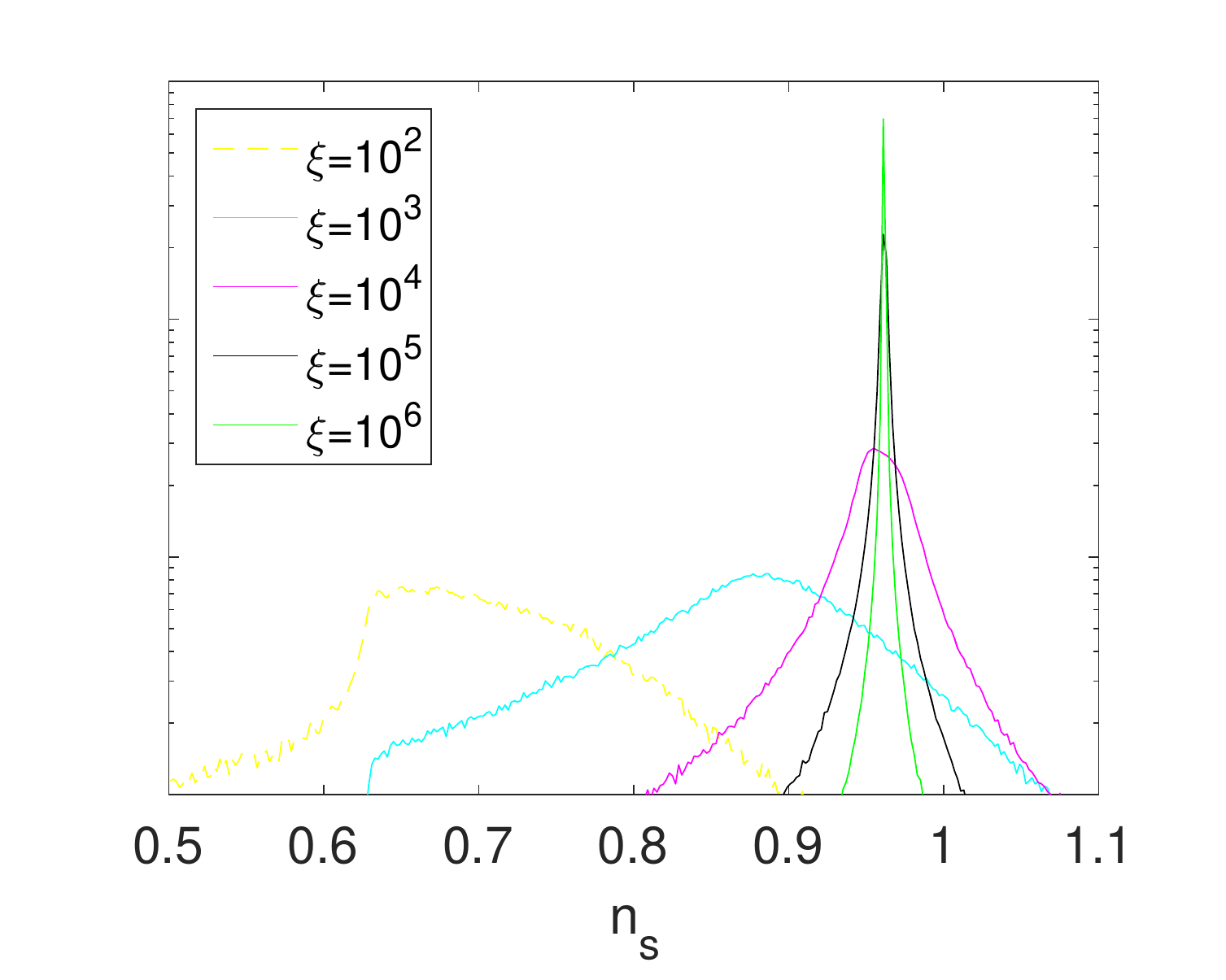}
\includegraphics[width=0.48\textwidth]{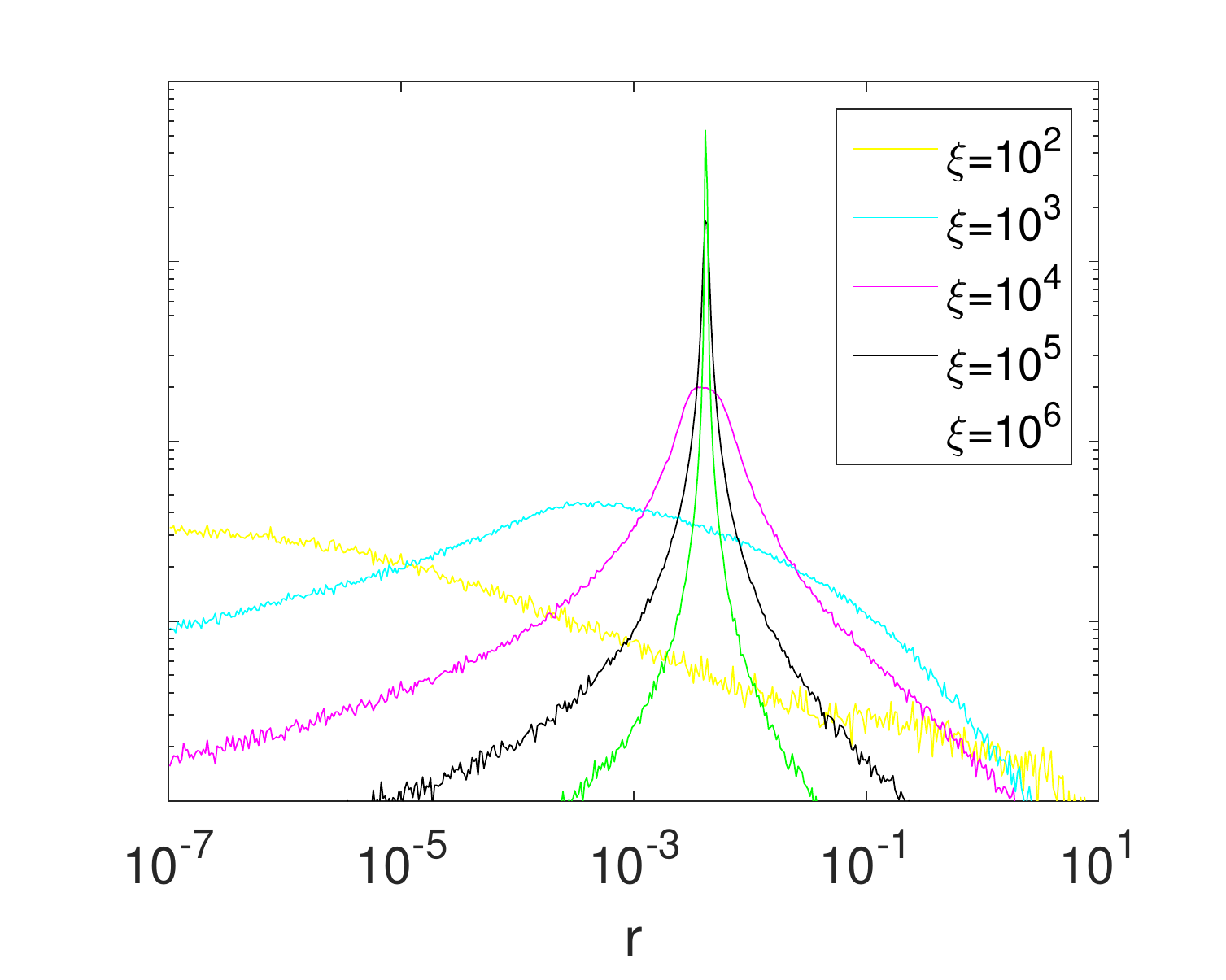}\\
\includegraphics[width=0.48\textwidth]{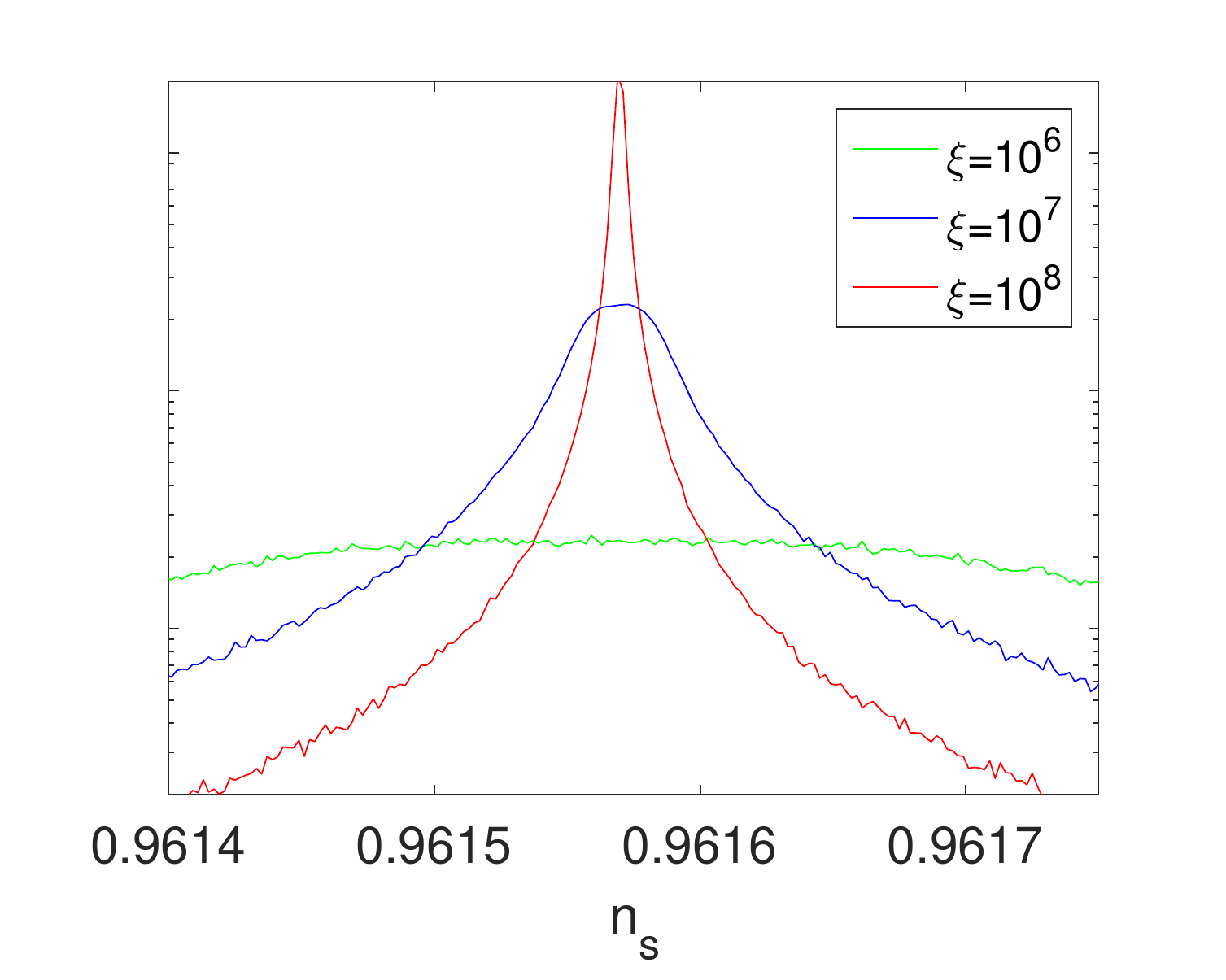}
\includegraphics[width=0.48\textwidth]{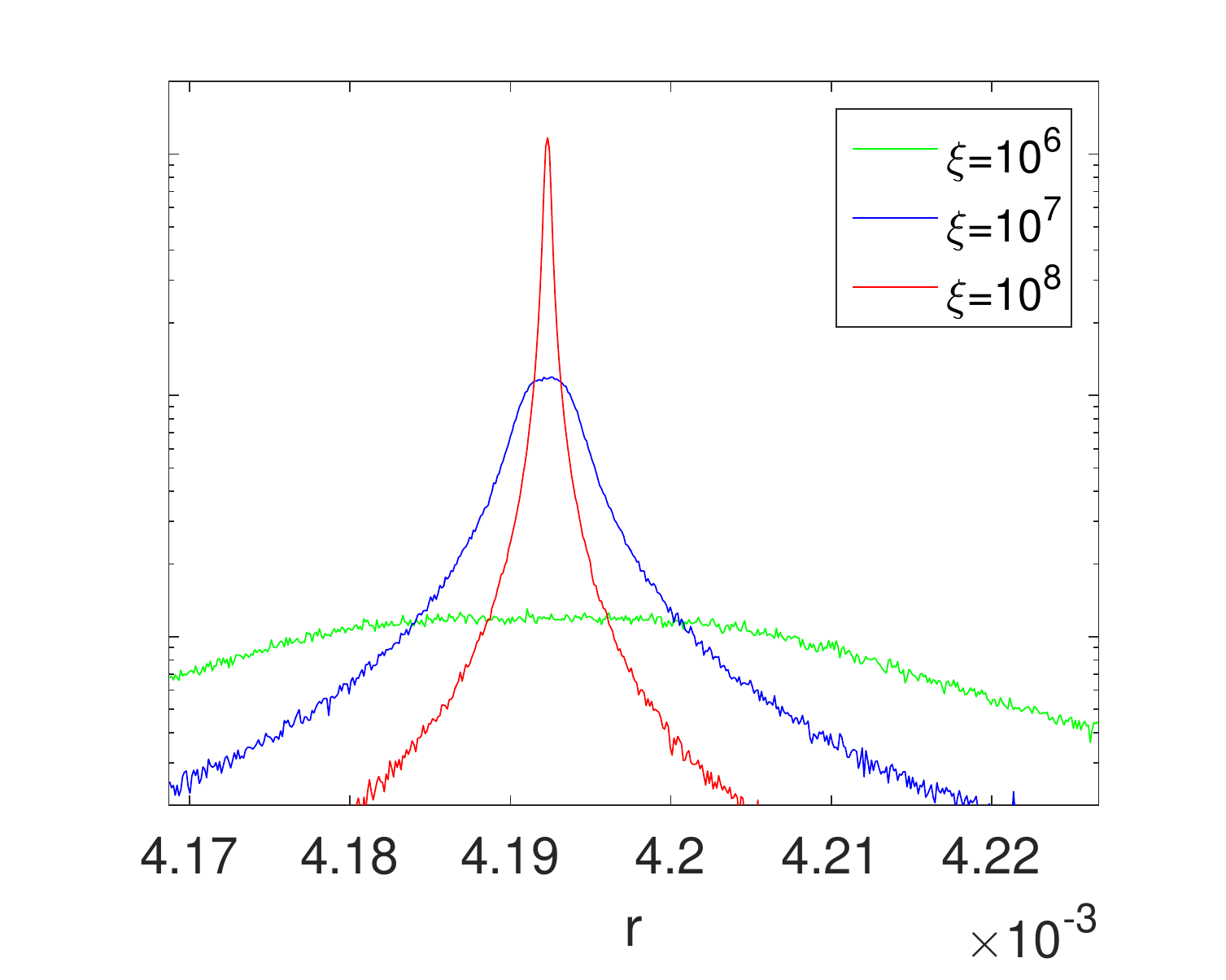}
\caption{\it Density profiles (on a log-scale) for different values of $\xi$. The left frames show the density profile for $n_s$, while the right frames show the density profile for $r$. The bottom frames are a zoom in around the Starobinsky point. Both $n_s$ and $r$ peak at the Starobinsky point for $\xi \gtrsim 10^4$.}
\label{fig:density_profiles}
\end{center}
\vspace{-0.5cm}
\end{figure*}

Turning to the numerical results, we start with a scatter plot in Fig.~\ref{fig:nsr_O1},  comparing the predictions for $\xi=10^2$ and $\xi=10^4$ with fixed $M_\Omega=5$, $M_V=10$ (and setting $N_e=50$). In perfect agreement with our analytic results, indeed a clearly visible line is present that goes from bottom left to top right through the Starobinsky point shown with a red star. Around this point, its slope is given by \eqref{slope}. Moreover, this line is much more pronounced for the larger value of $\xi$.

Studying models close to the Starobinsky point is difficult using scatter plots, since the finite point size blurs too much information regarding the density of points. Therefore, to be able to make any observation regarding the onset of the universal attractor regime, one should consider the density of the spectrum. For this we binned the data in small bins of $n_s$ ($r$) and counted the number of points within each bin, thereby marginalizing over $r$ ($n_s$). The resulting curve is a rough measure for the probability distribution of the variable, since the number of points over which is sampled is large. For a true measure of the probability, the spectrum has to be normalized. However, we only calculated the number of points in a bin, divided by the total number of points, which actually depends on the chosen binsize; fortunately, this will not influence our conclusions.

The density plots for $n_s$ and $r$ are shown in Fig.~\ref{fig:density_profiles}. In these plots it is clear that for $\xi=10^2$, the Starobinsky point is not of any importance, and the ensemble is most likely to be found in a hilltop state. When $\xi=10^4$ a peak is clearly visible at the Starobinsky point, and this peak sharpens when $\xi$ increases, just as the analysis in section \ref{retain} demonstrated. This centering around the Starobinsky point is a continuous process, starting from around $\xi\approx N_e^2$.

There is one final probe we want to present here that shows the emergence of the attractor phase, and that is the percentage of the number of \emph{non-trivial} outcomes of inflation. As explained before, a random model can have different outcomes of inflation, depending on the shape of the potential and the frame function. However, if the attractor phase is reached at infinite $\xi$, the outcome becomes independent of the model, and hence all models should be non-trivially ending. 
\begin{figure}[t!]
\centering
\includegraphics[scale=0.5]{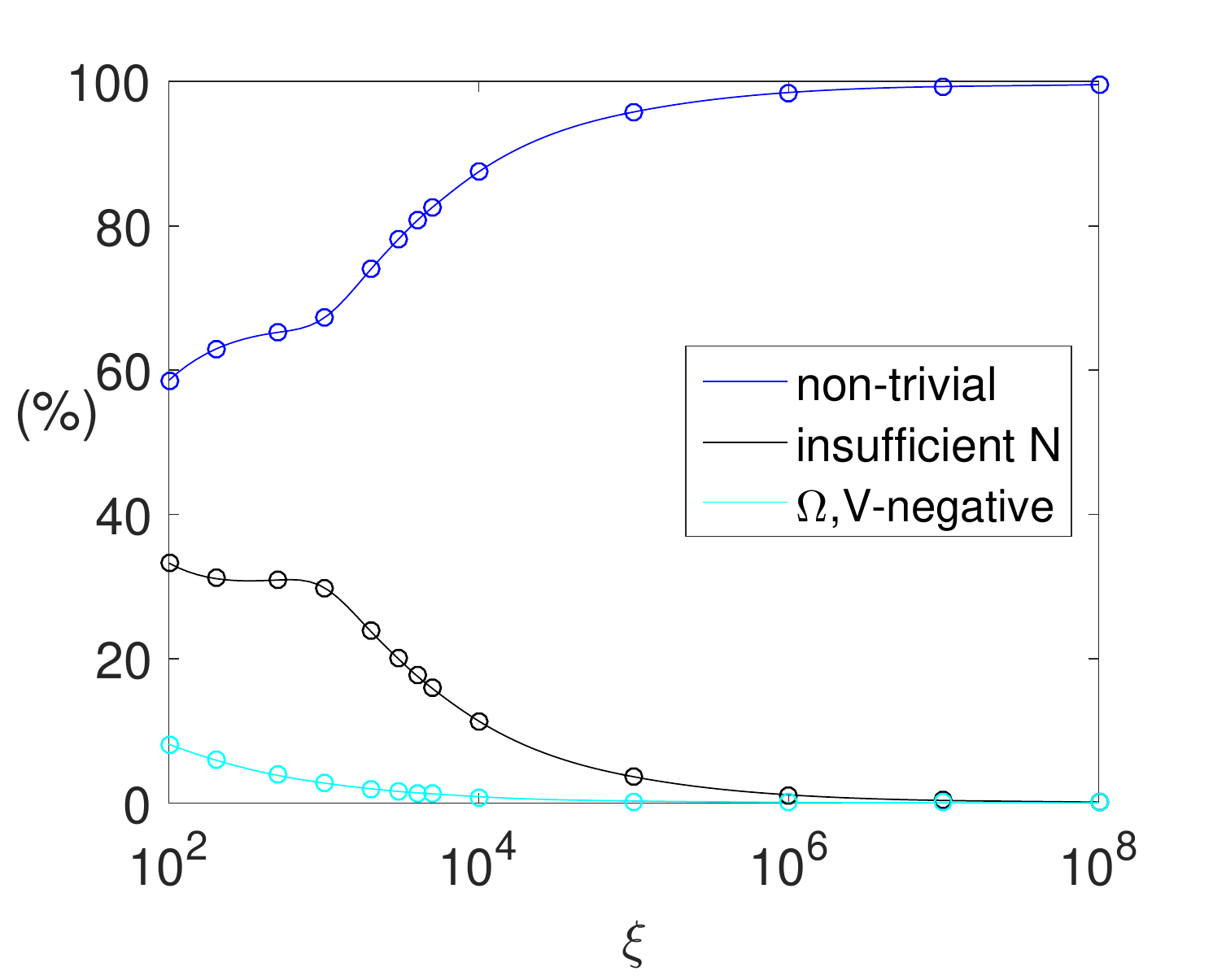}
\caption{\emph{The occurrence of different late-time behaviours as a function of $\xi$. The circles denote actual data points, the lines are only to guide the eye.}}
\label{fig:distribution}
\end{figure}
To probe this we plot the percentage of the number of outcomes in Fig.~\ref{fig:distribution}. The probability that a model ends non-trivially indeed increases when $\xi$ increases, and the number of models with insufficient e-folds to account for the observations (\emph{insuf}) and the number of models with negative potential and/or frame function during inflation (\emph{Vneg}) decrease. 

Note that in Fig.~\ref{fig:distribution} we observe the maximal increase of the number of non-trivial points around $\xi=10^4$. Also $\xi=10^4$ was the location where the peak was first centred around the Starobinsky point. We hence conclude that the lower bound $\xi\gtrsim 10^4$ appears first from CMB normalization arguments and our toy model analysis in subsection \ref{retain} and follows to be a special value also in the numerical study.

\section{Discussion}\label{discussion}

In this work, we have revisited non-minimally coupled inflation models in the spirit of \cite{Salopek:1988qh, Bezrukov:2007ep, Kallosh:2013tua}. Our interest was whether there exists a value of the non-minimally coupling strength that is preferred not only by matching COBE normalisation.

We first described how the non-minimal coupling $\xi$ may be used to induce an effective shift-symmetry which is protected against a possibly infinite tower of higher order corrections. The size of the non-minimal coupling determines the field range of this Einstein frame shift-symmetry. We identified two distinct regimes:
\begin{itemize}
 \item
 $\xi \sim \mathcal{O}(N^2_e)$: In this regime, the Jordan frame field is mostly sub-Planckian during inflation. As a consequence, it is inherently protected from most higher order terms, and may only be affected by a single correction term to the square relation \eqref{square}. Inflation will be driven by an intermediate plateau of hilltop potential generating at least $N_e$ e-folds. The inflationary predictions will therefore be roughly similar to those of PLANCK.
 \item
 $\xi > {O}(N^2_e) $: For larger values, the Jordan field only takes small values during inflation, and inflation is therefore protected from any higher-order term and is effectively governed the square relation \eqref{square}. Due to the larger non-minimal coupling, the intermediate plateau is prolonged such that the inflationary observables begin to converge towards the sweet spot of PLANCK. The predictions will have entered the 2-$\sigma$ contours of PLANCK once $\xi\sim\mathcal O(10^4)$. This lower bound is in remarkable agreement with the value of $\xi$ required to match the scalar perturbation amplitude $A_s$ \cite{Bennett:1996ce}.
 \end{itemize}
In the numerical component of this work, we parametrized non-minimal coupling functions and potentials as arbitrary polynomials. Drawing the  coefficients of the polynomials randomly, we examined the resulting Einstein frame potentials to find out whether observationally viable slow-roll inflation occurs. We found that with increasing non-minimal coupling $\xi$, the number of non-trivial inflationary trajectories increases. Remarkably, this increase is most pronounced in the range $\xi\sim\mathcal O(N_e^2)$ to $\xi\sim\mathcal O(10^4)$. Furthermore, we found that at $\xi\sim\mathcal O(N_e^2)$ there is a transition from a peak at low $n_s$ to a peak at the Starobinsky prediction of $n_s=0.962$. 

In other words, a non-minimal coupling $\xi$ can induce a shift-symmetry protected against all higher order terms (i.e.\ length of an inflationary plateau). The preferred value to match the COBE normalization coincides with the inflationary observables taking PLANCK-compatible values. 

To have a prediction of the implications of the assumption of factorial fall-off of the coefficients we repeated the analysis with choosing the random interval as $[-1,1]$ for $a_n, b_n$ in \eqref{general}. Though, as will be explained in appendix \ref{higherO}, the low order truncations of this system were different, the truncation independent regime showed the same observations. Thus we conclude that the above analysis is independent of the choice of the prior interval. Regarding the type of series used, for instance using Fourier series instead of polynomials, we expect that our main finding; that for large $\xi$ all models are located around the Starobinsky point, is still valid. However, the approach towards this point, i.e.~the predictions for $\xi\sim\mathcal O(N_e^2)$ and $\xi\sim\mathcal O(10^4)$, might in general be different as well as how these models approach the Starobinsky point, i.e.~Fig \ref{fig:nsr_O1}. Studying the model dependence of the predictions is an interesting follow-up analysis.
\acknowledgments

This work has been supported by the ERC Consolidator Grant STRINGFLATION under the HORIZON 2020 contract no.\ 647995 and by the German Science Foundation (DFG) within the Collaborative Research Center 676 "Particles, Strings and the Early Universe". BB was further supported by a travel grant of the PIER Helmholtz Graduate school and appreciates the RUG HEP theory group's hospitality during his stay. 
\appendix

\section{Inflationary Observables}\label{derivation}

We now outline how to derive the leading order terms of expressions \eqref{universal} and \eqref{spectralindex}. Consider the potential slow-roll parameters for a canonical inflaton $\chi$
\begin{equation}
\epsilon_V=\frac{1}{2}\left(\frac{1}{V}\frac{dV}{d\chi}\right)^2\thinspace ,\quad \text{and}\quad \eta_V=\frac{1}{V}\frac{d^2V}{d\chi^2}\thinspace .
\end{equation}
The number of e-folds $N_e$ is
\begin{equation}\label{aefolds}
N_e=\int\frac{1}{\sqrt{2\thinspace\epsilon_V}}d\chi\thinspace .
\end{equation}
Given Lagrangian \eqref{EinsteinL} with $\xi\gtrsim 1$ and
\begin{equation}
\Omega=1+\xi f(\phi)\thinspace , \quad V_J=\lambda f^2(\phi)\thinspace, 
\end{equation}
the inflationary potential in canonical fields is
\begin{equation}
V=V_0\left(1-e^{-\kappa\chi}\right)^2\thinspace,
\end{equation}
where $V_0=\lambda/\xi^2$ and $\kappa=\sqrt{2/3}$. Considering the potential and its derivatives to first order in $e^{-\kappa\chi}$, we may evaluate \eqref{aefolds} to obtain 
\begin{equation}\label{aresultn}
N_e=\frac{1}{2\thinspace\kappa^2}e^{\kappa\chi}\thinspace .
\end{equation}
Considering potential and derivatives only to leading order and recalling the expressions for the spectral index $n_s$ and the tensor-to-scalar ratio $r$, we substitute \eqref{aresultn} to obtain
\begin{equation}
n_s=1+2\eta_v-6\epsilon_v=1-\frac{2}{N}+\ldots\thinspace ,\quad \text{and}\quad r=16\epsilon_V=\frac{12}{N_e^2}+\ldots\thinspace ,
\end{equation}
where we have omitted the calculation of subleading terms as presented in \cite{Roest:2013fha}.

For our generic ansatz \eqref{general}, the canonical inflaton potential after conformal transformation is
\begin{equation}\label{VEplt1}
V_E=\frac{\lambda}{a_1^2 \xi^2}\left(1-\frac{1}{\Omega}\right)^2\left[b_2+\sum_{k=1}b_{k+2}\left(\frac{\Omega-1}{a_1 \xi}\right)^k\right]\thinspace .
\end{equation}
Recalling canonical normalisation \eqref{omega}, the above can be expanded to leading order during inflation as
\begin{equation}\label{risingap}
V=V_0\left(1-2e^{-\kappa\chi}+\xi^{-1}\frac{b_3}{b_2} e^{\kappa\chi}+\ldots\right)\thinspace .
\end{equation}
Expression \eqref{aefolds} can be evaluated exactly for the approximation given above, but its full form is lengthy. We hence point to the leading order terms given in \eqref{importantN}. It is straightforward to evaluate the potential slow-roll parameters with potential \eqref{risingap}. The crucial ingredient in order to predict the correct slope \eqref{slope} in the $n_s, r$ plot is to consider \eqref{importantN} up to order $\Omega^3$. Then, solving for $\Omega(N_e)$ is still analytically tractable.\footnote{The solution to a cubic equation may be complex. We choose the branch such that the resulting expressions for $n_s$ and $r$ are real.} Substituting the suitable solution into the evaluated slow-roll parameters then yields our main findings \eqref{spectralindex}.

\section{Higher Order Terms}\label{higherO}

The presented analysis has demonstrated that given $a_1, b_2, b_3 \sim\mathcal O(1)$ and $\xi > \mathcal O(N_{e}^2)$, inflation occurs with a leading order Starobinsky (or Hilltop) signature and a value of $\xi\gtrsim \mathcal{O} (10^4)$ can serve to push \emph{all} higher order corrections sufficiently far away in field space to arrive at an observationally viable model. We hence find an inflationary regime independent of the truncation of either series in \eqref{general}. 

However, due to the randomness of the coefficients $a_n, b_m$, it could in principle happen that terms $b_m\phi^m, m>2$ in the potential evade the $\xi$-induced flattening and influence the inflationary dynamics. Changing our set-up to $a_n, b_m \in [-1, 1]$, we now examine whether or not the set-up remains truncation independent when the coefficients are drawn such that terms $b_m\phi^m$ for $m>2$ are important, i.e.\ greater than unity, during inflation; in other words, the Jordan frame field $\phi$ is trans-Planckian to maintain the required amount of e-folds. Having the coefficients $a_n, b_m$ resemble a factorial suppression pattern, the non-canonical field has to be $\phi\gtrsim \mathcal O(M)$ during inflation ($M$ is the order of the frame function's truncation) for higher order terms to be non-negligible. Simply taking $a_n, b_m \in [-1, 1]$, the non-canonical field has to be $\phi\gtrsim\mathcal O(1)$ during inflation to feel the effect of higher order terms.

In what follows, we study the case $a_n, b_m \in [-1, 1]$ and $\phi\gtrsim\mathcal O(1)$ but the argument readily extends to the scenario $a_n, b_m \in [-1/n!, 1/n!]$ and $\phi\gtrsim\mathcal O(M)$. Consider
\begin{align} 
\Omega(\phi)=1+\xi \sum_{n=1}^M a_n\phi^n,\quad V_J(\phi)=\lambda \sum_{m=2}^{2M+\Delta}b_m\phi^m,
\end{align}
where $\Delta$ is a positive integer and hence parametrizes how much the highest order term of the Jordan frame potential departs from a square relation with the highest order term in the non-minimal coupling function $\Omega$. When $\phi>1$, we obtain  the effective potential 
\begin{equation}\label{HighestOrderN}
V_E\sim\frac{\lambda}{a_M^2\xi^2}\left[b_{2M}+\sum_{k=1}^{\Delta}b_{2M+k}\left(\frac{\Omega}{a_M\xi} \right)^{\frac{k}{M}}\right].
\end{equation}
If the potential departs from the square relation between potential and frame function at highest order, the Einstein frame potential in principle feels this effect. While also this effect can be made negligible by tuning $\Delta$ or simply pushing it away in field space by enlarging $\xi$, it could as such play an important role when the coefficients $b_m$ are drawn such that terms of the order $>2M$ become dominant in the inflationary region of the Einstein frame potential.
\begin{figure}[t!]
\centering
\includegraphics[scale=0.5]{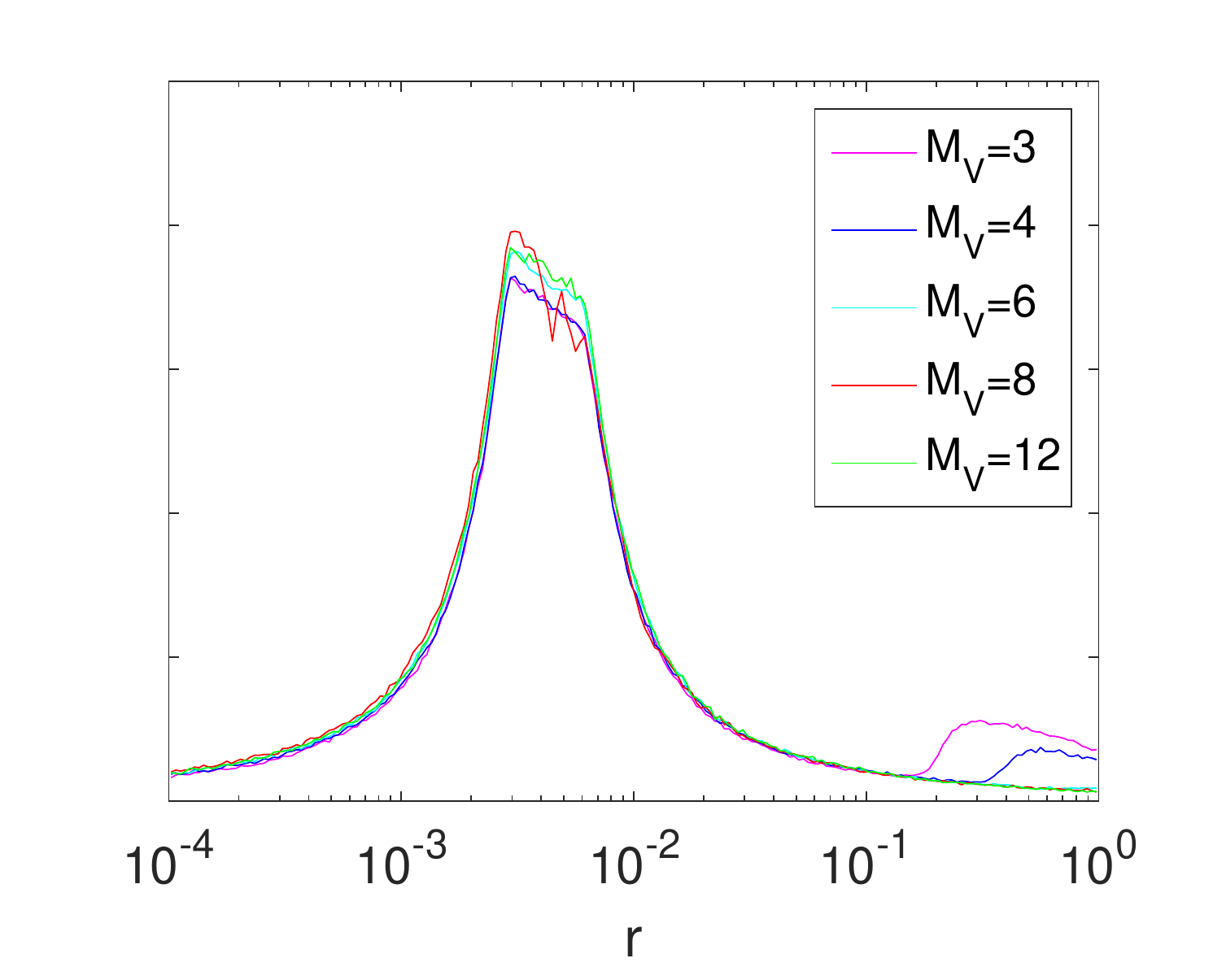}
\caption{\it Density profile for $r$ with $\xi=10^4$, $M_\Omega=1$ and with coefficients $b_m$ that are not factorially suppressed.}
\label{fig:distributionNonFac}
\end{figure}
As coefficients $b_{m>2M}$ may have either sign, the effect of these higher order terms on the inflationary dynamics can either be to curve the potential upwards and hence increase the number of chaotic signatures in the $n_s, r$ plot or to induce a hilltop and thus to enlarge the number of signatures with redder $n_s$ and very small $r$. We conjecture that a large $\Delta$ will increase the number of hilltop signatures while chaotic signatures may only be visible when $\Delta\sim\mathcal O(1)$ and $M$ is not too large. This is because a large $\Delta$ will allow for an interplay of coefficients $b_{m>2M}$ with possibly different signs such that hilltops occur whereas if there exists just one or two higher order terms, a positive highest order coefficient could be sufficient to steepen the potential before lower order terms will have induced a hilltop. The phenomenology of this analysis is depicted in figure \ref{fig:distributionNonFac}. This shows how chaotic signatures are only visible for $\Delta\sim\mathcal O(1)$.

We thus find that once sufficiently large $\xi\gtrsim\mathcal O(N_e^2)$ drives the non-canonical field displacement sub-Planckian, the form of the higher order coefficients is mostly irrelevant for the inflationary predictions.
\bibliography{numerical-attractor_v2}

\end{document}